%Paper: chao-dyn/9308006
%From: gallavotti%vaxrom.hepnet@Lbl.Gov
%Date: Sun, 22 Aug 93 07:16:44 PDT
%Date (revised): Sun, 22 Aug 93 07:31:00 PDT

%%% Plain TeX
%%%%%%%%%%%%%%%%%%%%%%%%%%%%%%%%%%%%%%%%%% FORMATO
\newcount\mgnf\newcount\tipi\newcount\tipoformule
\mgnf=0          %ingrandimento
\tipi=2          %uso caratteri: 2=cmcompleti, 1=cmparziali, 0=amparziali
\tipoformule=1   %=0 da numeroparagrafo.numeroformula; se no numero
                 %assoluto
%%%%%%%%%%%%%%%%%%%%%%%%%%%%%%%%%%%%%%%%%% INCIPIT
\ifnum\mgnf=0
   \magnification=\magstep0\hoffset=0.cm
   \voffset=-0.5truecm\hsize=16.5truecm\vsize=25.truecm
   \parindent=4.pt\fi
\ifnum\mgnf=1
   \magnification=\magstep1\hoffset=0.truecm
   \voffset=-0.5truecm\hsize=16.5truecm\vsize=24.truecm
   \baselineskip=14pt plus0.1pt minus0.1pt \parindent=12pt
   \lineskip=4pt\lineskiplimit=0.1pt      \parskip=0.1pt plus1pt\fi
%%%%%%%%%%%%%%%%%%%%%%%%%%%%%%%%%%%%%%%%%%
%%%%%GRECO%%%%%%%%%
%
   \let\g=\gamma     
  \let\h=\eta   \let\th=\vartheta \let\l=\lambda
               \let\p=\pi    
     \let\f=\varphi
  \let\o=\omega 
     
     \let\F=\Phi       

%%%%%%%%%%%%%%%%%%%%%%%%%%%%%%%%%%%%%%%%%%%%%%%%%%%%%%%%%%%%%%
%%%%%%%%%%%%%%%%%%%%%  NUMERAZIONE PAGINE

{\count255=\time\divide\count255 by 60 \xdef\oramin{\number\count255}
        \multiply\count255 by-60\advance\count255 by\time
   \xdef\oramin{\oramin:\ifnum\count255<10 0\fi\the\count255}}

\def\ora{\oramin }

\def\data{\number\day/\ifcase\month\or gennaio \or febbraio \or marzo \or
aprile \or maggio \or giugno \or luglio \or agosto \or settembre
\or ottobre \or novembre \or dicembre \fi/\number\year;\ \ora}

\setbox200\hbox{$\scriptscriptstyle \data $}

\newcount\pgn \pgn=1
\def\foglio{\number\numsec:\number\pgn
\global\advance\pgn by 1}
\def\foglioa{A\number\numsec:\number\pgn
\global\advance\pgn by 1}

%%%%%%%%%%%%%%%%% EQUAZIONI CON NOMI SIMBOLICI (brevetto Eckmann)
%%%%%%%%%%%%%%%%%%%%%%%%%%%%%%%%%%%%%%%%%%%%%%%%%%%%%%%%%%%%%%%%%

\global\newcount\numsec\global\newcount\numfor
\global\newcount\numfig
\gdef\profonditastruttura{\dp\strutbox}
\def\senondefinito#1{\expandafter\ifx\csname#1\endcsname\relax}
\def\SIA #1,#2,#3 {\senondefinito{#1#2}
\expandafter\xdef\csname #1#2\endcsname{#3} \else
\write16{???? ma #1,#2 e' gia' stato definito !!!!} \fi}
\def\etichetta(#1){(\veroparagrafo.\veraformula)
\SIA e,#1,(\veroparagrafo.\veraformula)
 \global\advance\numfor by 1
% \write15{\string\FU (#1){\equ(#1)}}
 \write16{ #1 => \equ(#1)  }}
\def \FU(#1)#2{\SIA fu,#1,#2 }
\def\etichettaa(#1){(A\veroparagrafo.\veraformula)
 \SIA e,#1,(A\veroparagrafo.\veraformula)
 \global\advance\numfor by 1
% \write15{\string\FU (#1){\equ(#1)}}
 \write16{ #1 => \equ(#1)  }}
\def\getichetta(#1){Fig. \verafigura
 \SIA e,#1,{\verafigura}
 \global\advance\numfig by 1
% \write15{\string\FU (#1){\equ(#1)}}
 \write16{ Fig. #1 => \equ(#1)  }}
\newdimen\gwidth
\def\BOZZA{
\def\alato(##1){
 {\vtop to \profonditastruttura{\baselineskip
 \profonditastruttura\vss
 \rlap{\kern-\hsize\kern-1.2truecm{$\scriptstyle##1$}}}}}
\def\galato(##1){ \gwidth=\hsize \divide\gwidth by 2
 {\vtop to \profonditastruttura{\baselineskip
 \profonditastruttura\vss
 \rlap{\kern-\gwidth\kern-1.2truecm{$\scriptstyle##1$}}}}}
\footline={\rlap{\hbox{\copy200}\ $\st[\number\pageno]$}\hss\tenrm
\foglio\hss} }
\def\alato(#1){}
\def\galato(#1){}
\def\veroparagrafo{\number\numsec}\def\veraformula{\number\numfor}
\def\verafigura{\number\numfig}
\def\geq(#1){\getichetta(#1)\galato(#1)}
\def\Eq(#1){\eqno{\etichetta(#1)\alato(#1)}}
\def\eq(#1){\etichetta(#1)\alato(#1)}
\def\Eqa(#1){\eqno{\etichettaa(#1)\alato(#1)}}
\def\eqa(#1){\etichettaa(#1)\alato(#1)}
\def\eqv(#1){\senondefinito{fu#1}$\clubsuit$#1
             \write16{#1 non e' definito!}%
\else\csname fu#1\endcsname\fi}
\def\equ(#1){\senondefinito{e#1}\eqv(#1)\else\csname e#1\endcsname\fi}

%%%%%%%%%
%\newcount\tipoformule
%\tipoformule=1   %=0 da numeroparagrafo.numeroformula; se no numero
%                 %assegnato
\ifnum\tipoformule=1\let\Eq=\eqno\def\eq{}\let\Eqa=\eqno\def\eqa{}
\def\equ{{}}\fi

\def\include#1{
\openin13=#1.aux \ifeof13 \relax \else
\input #1.aux \closein13 \fi}
\openin14=\jobname.aux \ifeof14 \relax \else
\input \jobname.aux \closein14 \fi
%\openout15=\jobname.aux %\write15
%
%%%%%%%%%%%%%%%%%%%%%%%%%%%%%%%%%%%%%%%%%%% CARATTERI %%%%%%%%%%%%%%%%
\newskip\ttglue
%%cm semplificato
\def\TIPI{
\font\ottorm=cmr8   \font\ottoi=cmmi8
\font\ottosy=cmsy8  \font\ottobf=cmbx8
\font\ottott=cmtt8  %\font\ottosl=cmsl8
\font\ottoit=cmti8
%%%%% cambiamento di formato%%%%%%
\def \ottopunti{\def\rm{\fam0\ottorm}% passaggio a tipi da 8-punti
\textfont0=\ottorm  \textfont1=\ottoi
\textfont2=\ottosy  \textfont3=\ottoit
\textfont4=\ottott
\textfont\itfam=\ottoit  \def\it{\fam\itfam\ottoit}%
\textfont\ttfam=\ottott  \def\tt{\fam\ttfam\ottott}%
\textfont\bffam=\ottobf
\normalbaselineskip=9pt\normalbaselines\rm}
\let\nota=\ottopunti}
%%%%%%%%
%%am
\def\TIPIO{
\font\setterm=amr7 %\font\settei=ammi7
\font\settesy=amsy7 \font\settebf=ambx7 %\font\setteit=amit7
%%%%% cambiamenti di formato %%%
\def \settepunti{\def\rm{\fam0\setterm}% passaggio a tipi da 7-punti
\textfont0=\setterm   %\textfont1=\settei
\textfont2=\settesy   %\textfont3=\setteit
%\textfont\itfam=\setteit  \def\it{\fam\itfam\setteit}
\textfont\bffam=\settebf  \def\bf{\fam\bffam\settebf}
\normalbaselineskip=9pt\normalbaselines\rm
}\let\nota=\settepunti}
%%%%%%%

%%cm completo
\def\TIPITOT{
\font\twelverm=cmr12
\font\twelvei=cmmi12
\font\twelvesy=cmsy10 scaled\magstep1
\font\twelveex=cmex10 scaled\magstep1
\font\twelveit=cmti12
\font\twelvett=cmtt12
\font\twelvebf=cmbx12
\font\twelvesl=cmsl12
\font\ninerm=cmr9
\font\ninesy=cmsy9
\font\eightrm=cmr8
\font\eighti=cmmi8
\font\eightsy=cmsy8
\font\eightbf=cmbx8
\font\eighttt=cmtt8
\font\eightsl=cmsl8
\font\eightit=cmti8
\font\sixrm=cmr6
\font\sixbf=cmbx6
\font\sixi=cmmi6
\font\sixsy=cmsy6
%%%%%%%%%%%%%%%%%%%%%%%%%%%%%%%%%%%%%%%
\font\twelvetruecmr=cmr10 scaled\magstep1
\font\twelvetruecmsy=cmsy10 scaled\magstep1
\font\tentruecmr=cmr10
\font\tentruecmsy=cmsy10
\font\eighttruecmr=cmr8
\font\eighttruecmsy=cmsy8
\font\seventruecmr=cmr7
\font\seventruecmsy=cmsy7
\font\sixtruecmr=cmr6
\font\sixtruecmsy=cmsy6
\font\fivetruecmr=cmr5
\font\fivetruecmsy=cmsy5
%%%% definizioni per 10pt %%%%%%%%
\textfont\truecmr=\tentruecmr
\scriptfont\truecmr=\seventruecmr
\scriptscriptfont\truecmr=\fivetruecmr
\textfont\truecmsy=\tentruecmsy
\scriptfont\truecmsy=\seventruecmsy
\scriptscriptfont\truecmr=\fivetruecmr
\scriptscriptfont\truecmsy=\fivetruecmsy
%%%%% cambio grandezza %%%%%%
\def \eightpoint{\def\rm{\fam0\eightrm}% switch to 8-point type
\textfont0=\eightrm \scriptfont0=\sixrm \scriptscriptfont0=\fiverm
\textfont1=\eighti \scriptfont1=\sixi   \scriptscriptfont1=\fivei
\textfont2=\eightsy \scriptfont2=\sixsy   \scriptscriptfont2=\fivesy
\textfont3=\tenex \scriptfont3=\tenex   \scriptscriptfont3=\tenex
\textfont\itfam=\eightit  \def\it{\fam\itfam\eightit}%
\textfont\slfam=\eightsl  \def\sl{\fam\slfam\eightsl}%
\textfont\ttfam=\eighttt  \def\tt{\fam\ttfam\eighttt}%
\textfont\bffam=\eightbf  \scriptfont\bffam=\sixbf
\scriptscriptfont\bffam=\fivebf  \def\bf{\fam\bffam\eightbf}%
\tt \ttglue=.5em plus.25em minus.15em
\setbox\strutbox=\hbox{\vrule height7pt depth2pt width0pt}%
\normalbaselineskip=9pt
\let\sc=\sixrm  \let\big=\eightbig  \normalbaselines\rm
\textfont\truecmr=\eighttruecmr
\scriptfont\truecmr=\sixtruecmr
\scriptscriptfont\truecmr=\fivetruecmr
\textfont\truecmsy=\eighttruecmsy
\scriptfont\truecmsy=\sixtruecmsy
}\let\nota=\eightpoint}

\newfam\msbfam   %per uso in \TIPITOT
\newfam\truecmr  %per uso in \TIPITOT
\newfam\truecmsy %per uso in \TIPITOT
%%%%%%%%%%%%%%%%%%%%%%%%%%%%%%%
%%%%%%%%%%%%%%%%%%%%%%%%%%%%%%%  Scelta dei caratteri
%\newcount\tipi \tipi=???? ===> e' definito all'inizio
\newskip\ttglue
\ifnum\tipi=0\TIPIO \else\ifnum\tipi=1 \TIPI\else \TIPITOT\fi\fi
%%%%%%%%%%%%%%%%%%%%%%%%
%%%%%%%%%%%%%%%%%%%%%%%%%%%%%%%%%%%%%%%%%% DEFINIZIONI VARIE
%
\def\V#1{\vec#1}\let\ciao=\bye
\let\ig=\int
 \let\0=\noindent

\def\guida{\leaders\hbox to 1em{\hss.\hss}\hfill}
\def\tende#1{\vtop{\ialign{##\crcr\rightarrowfill\crcr
              \noalign{\kern-1pt\nointerlineskip}
              \hglue3.pt${\scriptstyle #1}$\hglue3.pt\crcr}}}
\def\otto{{\kern-1.truept\leftarrow\kern-5.truept\to\kern-1.truept}}

%%%%%%%%%%%%%%%%%%%%%%%%%%%%%%%%%%%%%%%%%%LATINORUM

\def\ie{\hbox{\it i.e.\ }}

\def\fiat{{}}

%%%%%%%%%%%%%%%%%%%%%%%%%%%%%%%%%%%%%%%%%%%
%%%%%%%%%%%%%%%%%%%%%%%%%%%%%%%%%%%%%%%%%%%DEFINIZIONI LOCALI

\def\={{ \; \equiv \; }}

\def\st{\scriptscriptstyle}
\def\fra#1#2{{#1\over#2}}
\let\0=\noindent
\def\*{\vskip0.3truecm}

\def\AA{{\V A}}\def\LL{{\cal L}}\def\EE{{\cal E}}\def\8{\hfill\break}
\def\eh{{e,\h}}

%%%%%%%%%%%%%%%%%%%%%%%%%%%%%%%%%%%%%%%%%%%%%%%%%%%%%%%%%
%\def\equ{}
%\BOZZA %\overfullrule=10pt
%

%\input fiat
\fiat
\centerline{ ROTATION AXIS VARIATION DUE TO SPIN ORBIT
RESONANCE}
\*
\centerline{ Giovanni Gallavotti \footnote{${}^*$}{\nota A
lecture given at the {\it $I^o$ Congresso Nazionale di Meccanica Celeste},
L'Aquila, may 22-- 26, 1993.  This text is archived in
$mp\_arc@math.utexas.edu$, \# 93-222; \TeX copies can also be obtained by
e-mail from: $gallavotti@vaxrom.infn.it$} }
\centerline{ Dipartimento di
Fisica, Universit\`a La Sapienza}
\centerline{ P.le Moro 2, 00185, Roma,
Italia}
\*
\0{\it keywords: planetary precession, rigid body, chaos, KAM,
Arnold diffusion, averaging, celestial mechanics, classical mechanics, large
deviations}
\*\*
\numsec=1\numfor=1\penalty10000

Let $\EE$ be a planet modeled by a homogeneous rigid ellipsoid with
symmetry axis $NS$, with polar inertia moment $J_3$, equatorial
moment $J$ and mechanical flattening coefficient $\h=(J_3-J)/J_3$.
\footnote{${}^1$}{\nota this means that if $R$ is the equatorial radius
then the polar radius is $R/(1+2\h)^{1/2}$.}

The planet center of mass $T$ is supposed to revolve on a keplerian orbit
$t\to\V r_T(t)$ about a focus $S$: the orbit plane will be called the
{\it ecliptic} plane and $\V{{\bar k}}$ will denote its unit normal
vector ({\it celestial north}) which sees the planet revolving
counterclockwise.

The longitude $\l_T$ of $\V r_T$ on the ecliptic will be reckoned from
the aphelion of the ellipse; hence $\l_T=0$ is the {\it aphelion}
position, \ie where $r_T \equiv |\V r_T|$ is maximal: $r_T(0) = a (1
+e)$, $a$ being the major semiaxis of the keplerian ellipse and $e$ its
eccentricity.

With the above conventions, $r_T$ and $\l_T$ are related by:
$r_T\=|\V r_T|=p\cdot(1-e\cos\l_T)^{-1},\ p\=a(1-e^2)$
and the Kepler's laws imply that if $\l$ is the keplerian {\it average
anomaly} then: $\l\=\l_T+2e\sin\l_T+\fra34e^2\sin2\l_T+\ldots$.
and the motion is $\l\to\l+\o_Tt$, where $2\p/\o_T = 2\p a^{3/2}
g_N^{-1/2}$ is the year of the planet, $g_N\=k(m_S+m_T)$ if $k$ is Newton's
constant and $m_T,m_S$ are the masses of the planet and of its star.

The planet is described by means of the following coordinates:\8
1) The total angular momentum, or {\it spin}, $M$ and its projections
$K\=M\cos i$ on the axis orthogonal to the ecliptic ("celestial
north--south axis").\8
2) The angles $\g,\f,\l$ where $\g$ is the angle between an axis fixed
on the ecliptic ({\it Aries line}) and the spin--ecliptic node (\ie the
intersection of the ecliptic and the plane orthogonal to the spin:
{\it equinox line}); $\f$ is the angle between the spin ecliptic node
and the analogous spin--equator node; and $\l$ is the average anomaly
of the planet, which rotates uniformly at angular velocity $\o_T$.\8
It is a well known theorem in classical mechanics (Andoyer--Deprit) that
the pairs $(M,\f)$ and $(K,\g)$ are canonically conjugate variables
(see [G], p.318) and, introducing an auxiliary variable $B$ canonically
conjugate to $\l$, the energy of the system can be written:
$$H=\o_T B+\fra{M^2}{2J}+ V=\o_T B+\fra{M^2}{2J}
-\ig_\EE{k m_T m_S\over|\V r_T+\V x|}{d\V x\over|\EE|}\Eq(1)$$
One could avoid introducing $B$: but one would then have a time
dependent hamiltonian, which I do not like.

The complete description of the rigid body configuration would require
an extra pair of canonically conjugated varibles, namely $(L,\psi)$
where $\psi$ is the angle between a comoving axis fixed on the equator
and the spin--equator node and $L$ is the projection $L=M\cos\th$ of
the spin on the $NS$ axis.  However, by symmetry, the hamiltonian does
not depend on $\psi$; hence $L$ is a constant of motion and one can see
that it only gives an additive contribution to the energy, which is
dropped in \equ(1).

I shall consider the following approximation for $V$:\8
a) the integral is developed in powers of the ratio $R/a$ and the
expansion will be truncated to the order $2$ included neglecting
the orders $4$ and higher.
\footnote{${}^2$}{\nota  only the 2d order is non trivial among the
first three orders.}\8
b) the expression thus obtained will be developed in powers of the
eccentricity and the expansion will be truncated to 2d order.

The model thus obtained will be called the {\it D' Alembert
precession-nutation model}.  The $e=0$ model was in fact used by
D'Alembert, [L], to deduce his celebrated theory of the equinox
precession for the Earth and the theory of the lunar plane precession.

It is not interesting to limit ourselves to the D'Alembert's case $e=0$
because the phenomenon of large variations of the inclination axis are
in this approximation not possible.

Define $\o_D\=M/J$, calling it conventionally the {\it daily
rotation}, and consider the line $\LL$ in the plane of the action
variables $\AA=(M,K)$ defined by:
$$M=J\o_D\=2 J\o_T=const,\qquad K_0\le K\le K_1\Eq(2)$$
and call $\AA^0$ and $\AA^1$ its extremes. Since $K=M\cos i$ one realizes
that proceeding on the above line from $\AA^0$ to $\AA^1$ the spin
inclination angle changes between $i_0$ and $i_1$ and the system is
locked in a $1:2$ spin orbit resonance.

A motion in which the projection in the $(M,K)$ plane closely follows
the line $\LL$ is therefore a motion during which a variation of the
inclination axis is observed. Along such line, which will be fixed in
the following, the ratio $\cos\th\=L/M$ is a constant which will
therefore also be fixed: and which will be supposed different from
$\pm1$, (\ie the angle $\th$, {\it nutation constant}, between the spin
and the $NS$ axis is supposed different from $0$ or $\p$).

The eccentricity will be taken $e=\h^c$ for some $c>1$ so that setting
$\h$ small implies that the eccentricity is also (much) smaller.

The question that will be investigated, see [CG], is whether there exists a
trajectory $X_\eh(t)$ starting at $t=0$ in $\AA_\eh=(M_\eh,K_\eh)$ near
$\AA^0$, with phases $(\g_\eh,\f_\eh,\l_\eh)$, and reaching after a
long enough time the vicinity of $\AA^1$.

If such a trajectory exists for all $\h$ small enough, {\it but
different from $0$}, one says that there is a {\it Arnold
diffusion} phenomenon.
\footnote{${}^3$}{\nota this refers to one possible definition of Arnold's
diffusion, [A]: it is a special case of a general definition
proposed in [CG].  But there are other definitions: different and,
often, not very precise.}

What is remarkable, when the diffusion phenomenon happens, is that it
happens in spite of the fact that $K,M$ are adiabatic invariants, \ie
{\it they are variables which in any {\sl averaging approximation} could
only change by quantities which are infinitesimal as $\eh\to0$}: while
in the above situation $K$ or $i$ change by a quantity $\sim i_1-i_0$
{\it independent on $\eh$}.

The existence of Arnold diffusion says therefore that no averaging
approximation can be even approximately correct over very long time
scales or for infinite time.

This is not at all a knell for the widely used averaging methods: but
it shows the interest of estimating the time scales over which the
phenomenon happens: it is becoming well understood that the correctness
of the averaging methods give very accurate results over time scales
that are often of the order of the age of the universe, [CeG].

The result, [CG], on the above question, in the case of the described
precession nutation model is the following.
\*
\0{\bf Theorem:} {\it Suppose that $[i_0,i_1]$ does not contain the
values $0,\fra\p2,\p$ and suppose that the angle $\th$ between the $NS$
axis and the spin axis does not vanish in the configurations in $\LL$
line.
\footnote{${}^4$}{\nota\rm it has been remarked above that the angle
$\th$ is constant on the line $\LL$.}
There are constants $c,b,d$ such that if $e=\h^c$ then
for all $\h$ small enough, but non zero, there are phases
$\F_\h=(\g_\h,\f_\h,\l_\h)$ and actions $\AA_\h=(M_\h,K_\h)$ such
that:\8
1) the trajectory $X_\h(t)=(\AA(t),\F(t))$, starting in the
configuration $\AA(0)=\AA_\h,\F(0)=\F_\h$ stays close, in the
$\AA$--variables, within $\h^b$ to
the line $\LL$ during a suitable time $\ge T_\h$.\8
2) the time $T_\h$ is such that $\AA(T_\h)$ is close (\ie within $\h^b$)
to $\AA^1$ while $\AA(0)$ is close to $\AA^0$, (\ie within $\h^b$).}
\*
Therefore in the above precession problem Arnold's diffusion is
possible.  The angle of spin inclination can change by an order of
magnitude $O(1)$ in a long enough time, no matter how small is the spin
orbit coupling (\ie $\h$), provided it is non zero and provided the
eccentricity is small enough compared to the flattening.  And it
becomes interesting to get an idea of the size of the time $T_\h$.

In [CG] an explicit estimate for $T_\h$ is derived:
\*
{\bf Theorem:} {\it in the situation of the previous theorem there is a
constant $f$ such that for $\h$ small enough the diffusion time $T_\h$
can be estimated to be not longer than:
$$\bar T_\h=\o_T^{-1} e^{(e^2\h^{3/2})^{-2}g(\LL)^{-2}\,f},\kern1cm
g(\LL)=\sin\th\,\min_{i_0\le i\le i_1}
\Big(\fra{1-cos i}{1+\cos i}\cos^2 i\,\sin i\Big)\Eq(3)$$
}

Hence the estimate is a very very long time, compared to the times over
which one expects the averaging methods to have some validity, which
have scale of order $\sim\exp f\h^{-1/2}$. Also the constants $b,c,d,f$,
whose values can actually be computed, see [CG], turn out to be very
poor for any practical application.

But to get to practical applications a lot of estimates would have to
be refined and it is not completely obvious that this is really
impossible.  One can recall, on this respect, that for a long time it
was stated by some people that the KAM estimates were too bad for any
practical use: and this turned out to be grossly incorrect,
[CeG],[CC],[LR].
In the case of diffusion the situation looks much harder, perhaps
desperately so: but we shall see.

This shows that the estimate of the time scale for the diffusion, \ie
essentially the time one has to wait to see a violation of
the predictions of the averaging approximations, diverges if one tries
to bridge gaps in inclination containing $i=0,\fra\p2,\p$.

In particular the above estimate diverges if one tries to find a
trajectory in which the spin sign is reversed (\ie $i$ goes through
$\fra\p2$). The difficulty in constructing such trajectories might be a
manifestation of a physical phenomenon and not just a defect of the
techniques of proof; it seems, indeed, that the chaotic motions of the
planetary axes are unable, in absence of dissipative effects, to change
the spin sign, \ie the sign of $K$, [LRo].

Formula \equ(3) is suggestive as it leads to the conjecture that
(in general) diffusion along a path $\LL$ in action space might really
take place over a time scale dependinng exponentially on some power of the
coupling constant {\it times a coefficient determined my maximising
a suitable functional defined on $\LL$}: therefore I call \equ(3),
by analogy, a {\it large deviation formula}.

I conclude with a technical comment: the analysis is based on the fact
that the hamiltonian system under consideration has three very
different time scales: namely the daily time scale ($\o_D^{-1}$,
coinciding with the year time scale because the free system is in a
$1:2$ resonance), the resonance time scale of order
$O(\h^{-1/2}\o_D^{-1})$ (describing the characteristic time of
oscillation transversal to the resonance) and the equinox precession
scale of order $O(\h^{-1}\o_D^{-1})$ (arising from an application of the
method of averaging and describing the precession of the equinox line,
when $e=0$).  The three time scales become very different from each
other when $\h\to0$: this is due to the degeneracy inherent in the
problem where the unperturbed hamiltonian {\it does not depend on the
variable $K$}. And it implies the phenomenon of {\it large angles at
the homoclinic intersections}, see [CG], which in turn implies the
existence of diffusion.  Deeply different is what happens near
resonances of non degenerate systems, where all the action variables
appear non linearly in the unperturbed hamiltonian: in such cases there
are only two distinct time scales describing the motions near the
resonances and the theory of diffusion is much harder (because the
homoclinic angles are exponentially small, see [CG], [G2]).

Thus the above theory is only one more instance of the (well known)
fact that the degeneration intrinsically present in all celestial
mechanics problems is very often, in fact, making the theory easier
rather than harder (as sometimes naively claimed).

For connections between the techniques used in the theory of the above
problem and the KAM theorem see also [G2],[G3].

\penalty-200

\vskip0.5cm
{\bf References:}
\penalty10000\*\penalty10000

\item{[A] } Arnold, V.: {\it Instability of dynamical systems with several
degrees of freedom}, Sov. Mathematical Dokl., 5, 581-585, 1966.

\item{[CC] } Celletti, A., Chierchia, L.: {\it Construction of analytic
*KAM surfaces and effective stability bounds}, Communications in
Mathematical Physics, 118, 119-- 161, 1988.

\item{[CG] } Chierchia, L., Gallavotti, G.: {\it Smooth prime integrals for
quasi-integrable Hamiltonian systems} Il Nuovo Cimento, 67 B, 277-295,
1982.

\item{[CeG] } Celletti, A., Giorgilli, A.: {\it On the stability of the
lagrangian points in the spatial restricted problem of three bodies},
Celestial Mechanics, 50, 31-- 58, 1991

\item{[G] } Gallavotti, G.: {\it The elements of Mechanics}, Springer, 1983.

\item{[G2] } Gallavotti, G.: {\it Twistless KAM tori, quasi flat
homoclinic intersections, and other cancellations in the perturbation
series of certain completely integrable hamiltonian systems.  A
review.}, archived in the electronic archive {\tt
mp\_arc@math.utexas.edu}, \#93-164; e-mail (last version) copies
in \TeX (plain of course) available also by direct request to the author.

\item{[G3] } Gallavotti, G.: {\it Twistless KAM tori},
archived in {\tt mp\_arc@math.utexas.edu} \#93-172; e-mail copies
(last version)
in \TeX (plain of course) available also by direct request to the author.

\item{[L] } de la Place, S.: {\it M\'ecanique C\'eleste}, tome II, book 5,
ch. I, 1799, english translation by Bodwitch, E., reprinted by Chelsea,
1966.

\item {[LR] } Llave, R., Rana, D.: {\it Accurate strategies for small
divisor problems}, Bullettin A.M.S., 22, 85-- 90, 1990. See also: {\it
Accurate strategies for K.~A.~M. bounds  and their implementation} , in
"Computer Aided proofs in Analysis", ed. K. Meyer, D. Schmidt, Springer
Verlag, 1991.

\item{[LRo] } Laskar, J. Robutel, P.: {\it The chaotic obliquity of the
planets}, Nature, 361, 608-- 612, 1993. See also the report of J. Laskar
at this conference.

\ciao